\documentclass[conference]{IEEEtran}
\usepackage{cite}
\usepackage{amsmath,amssymb,amsfonts}
\usepackage{algorithm}
\usepackage{algpseudocode}
\usepackage{microtype}
\usepackage{graphicx}
\usepackage{textcomp}
\usepackage{xcolor}
\usepackage{hyperref}
\usepackage{float}
\def\BibTeX{{\rm B\kern-.05em{\sc i\kern-.025em b}\kern-.08em
    T\kern-.1667em\lower.7ex\hbox{E}\kern-.125emX}}

\makeatletter
\newcommand{\linebreakand}{%
  \end{@IEEEauthorhalign}
  \hfill\mbox{}\par
  \mbox{}\hfill\begin{@IEEEauthorhalign}
}
\makeatother

\begin{document}

\title{EDMFormer: Genre-Specific Self-Supervised Learning for Music Structure Segmentation}

\author{

\IEEEauthorblockN{Sahal Sajeer}
\IEEEauthorblockA{
    \textit{University of Waterloo} \\
    ssajeer@uwaterloo.ca
}

\and
\IEEEauthorblockN{Krish Patel}
\IEEEauthorblockA{
    \textit{University of Waterloo} \\
    k77patel@uwaterloo.ca
}

\and
\IEEEauthorblockN{Oscar Chung}
\IEEEauthorblockA{
    \textit{University of Waterloo} \\
    ohchung@uwaterloo.ca
}

\and
\IEEEauthorblockN{Joel Song Bae}
\IEEEauthorblockA{
    \textit{University of Waterloo} \\
    js2bae@uwaterloo.ca
}


%
%

} 
\maketitle

\begin{abstract}
Music structure segmentation is a key task in audio analysis, but existing models perform poorly on Electronic Dance Music (EDM). This problem exists because most approaches rely on lyrical or harmonic similarity, which works well for pop music but not for EDM. EDM structure is instead defined by changes in energy, rhythm, and timbre, with different sections such as buildup, drop, and breakdown. We introduce \textbf{EDMFormer}, a transformer model that combines self-supervised audio embeddings using an EDM-specific dataset and taxonomy. We release this dataset as \textbf{EDM-98}: a group of 98 professionally annotated EDM tracks. EDMFormer improves boundary detection and section labelling compared to existing models, particularly for drops and buildups. The results suggest that combining learned representations with genre-specific data and structural priors is effective for EDM and could be applied to other specialized music genres or broader audio domains.
\end{abstract}

\section{Introduction}

Music structure analysis (MSA) is a core task in music information retrieval (MIR), enabling applications like automatic song summarization, playlist generation, DJ assistance, and music recommendation \cite{paulus2010state}. Modern MSA systems often rely on self-supervised learning (SSL) to create audio embeddings from large collections of music \cite{liu2022audio}. Transformer-based models such as SongFormer perform well by learning harmonic and melodic similarity patterns across time \cite{songformer2025}. 

However, these models assume that structure is mainly defined by harmonic repetition and lyrical phrasing, such as in Western pop music (e.g., verse, chorus, bridge). This assumption fails for EDM, where structure is driven by different features. EDM sections such as buildup and breakdown are not characterized by chords or vocals, but by spectral brightness, rhythmic density, and low-frequency energy changes \cite{poumtchak2010}. As a result, even top-tier MSA models often mislabel EDM sections, revealing a systematic genre mismatch.

\subsection{Motivation}

Prior MIR research shows that genre mismatches reduce model performance when learned representations do not match the structural patterns of the target genre \cite{joung2024domain}. For EDM, switch point detection (SPD) studies indicate that energy novelty, drum onset counts, and timbral features are the best cues for structural transitions \cite{spd2024}. These cues are mostly absent in the current SSL embeddings trained on pop music. It is also noted that rhythmic and energy structuring can improve boundary detection by over 3\% when combined with learned embeddings \cite{spd2024}.

These observations suggest that SSL-based MSA models miss important information needed for energy-driven music due to an inadequate taxonomy and lack of relevant data. This finding motivates an EDM-specific approach that integrates a more appropriately structured taxonomy and data.

\subsection{Related Works}

Early MSA approaches relied on handcrafted features such as chroma and novelty curves to detect structural boundaries \cite{paulus2010state}. With the rise of deep learning, convolutional and recurrent neural networks replaced feature engineering with learned representations. More recently, however, transformer-based models train with SSL objectives, such as MuQ and MusicFM, enabling models like SongFormer to achieve high performance on pop music datasets \cite{muq2025, won2023musicfm, songformer2025}.

However, improved representation learning does not necessarily translate to optimal structural segmentation. Several studies report that SSL-only embeddings underperform when domain-specialized rhythmic cues are ambiguous. Beat-focused feature research shows that explicit rhythmic taxonomy setups can significantly improve boundary detection performance, particularly when fused with neural embeddings \cite{spd2024}. Similarly, Joung et al. demonstrate that domain specialization improves model robustness in MIR tasks \cite{joung2024domain}.

For EDM, SPD research identifies energy novelty, drum onset counts, and spectral flux as the strongest indicators of structural change \cite{spd2024}. But no existing MSA work has incorporated these findings into a domain-specific structure segmentation taxonomy. Additionally, existing datasets such as SALAMI still contain minimal EDM representation and use pop-centric taxonomies \cite{salami2011}.

These gaps motivate the need for (1) an EDM dataset, (2) an EDM-specific structural taxonomy, and (3) a SSL transformer model that combines the two.

\subsection{Problem Definition}

Given an audio waveform $X$, the goal of music structure segmentation is to predict a sequence of timestamps 
$T = \{t_1, t_2, ..., t_n\}$ representing structural boundaries, along with corresponding section labels 
$Y = \{y_1, y_2, ..., y_n\}$.

Existing approaches typically learn a general mapping
\begin{equation} \label{f_general}
f_{general}: E_{ssl}(X) \xrightarrow{\text{pop-centric dataset + general taxonomy}} (T, Y),
\end{equation}
where $E_{ssl}(X)$ are embeddings extracted from pretrained self-supervised music models, trained on diverse music corpora.

In contrast, our approach defines a domain-specialized mapping
\begin{equation} \label{f_edm}
f_{edm}: E_{ssl}(X) \xrightarrow{\text{EDM-98 dataset + EDM taxonomy}} (T, Y),
\end{equation}
where the same SSL embeddings are used, but training leverages the EDM-98 dataset and its EDM-specific structural taxonomy, allowing the model to learn accurate patterns and section labels.

\section{Methodology}

\subsection{EDM-98 Dataset}

We curated EDM-98, a dataset of 98 EDM tracks across five tempo categories (120-180 BPM) using its source corpus EDM-CUE, which contains 4710 songs \cite{edmcue2024}. Two professional annotators labelled structural sections (intro, build-up, drop, breakdown, outro, silence, and end) with ±0.5s precision. The dataset was constructed as a stratified, equal‑allocation subset to counter the heavy tempo imbalance of EDM‑CUE.

For our implementation, each track's BPM was extracted (with tracks lacking BPM excluded) and then assigned to fixed buckets. The sample allocation was distributed as evenly as possible across these buckets, with residual capacity handling to reach N=98. Selection within each stratum was randomized without replacement.

The resulting EDM‑98 marginal tempo distribution was intentionally close to uniform across mid/high‑tempo strata (House/Deep House 19.4\%; Electro/Dance, Techno, Trance/Progressive, Dubstep/Bass, Drum \& Bass/Jungle each 14.3\%; Chill/Downtempo 9.2\%). This distribution produced a broad BPM spread (mean 137.31, median 135.00, SD 20.04, range 100-175).

In contrast, the full EDM‑CUE dataset is highly skewed toward high‑tempo Drum \& Bass/Jungle, with the train split (N=4710) showing BPM with mean 169.52, median 174.0, and SD 11.62. 90.5\% of all tracks are this subgenre \cite{edmcue2024}. This severe class imbalance yields a markedly different marginal tempo distribution than EDM-98, which intentionally flattens the tempo histogram to support robust model training and evaluation across a wider tempo/genre spectrum. Thus, EDM‑98 should be interpreted as a balanced, purpose‑built subset rather than a proportional microcosm of EDM‑CUE.

\subsection{EDM-Specific Taxonomy}

All tracks in EDM-98 were labelled using an EDM-specific taxonomy designed to reflect changes in energy. As traditional pop-oriented taxonomies (verse, chorus, bridge) failed to capture EDM’s structure, we defined a new EDM taxonomy with the following categories:

\begin{itemize}
    \item \textbf{Intro:} Beginning of the track, low energy, sparse instrumentation.
    \item \textbf{Build-up:} Gradually increasing energy and tension, often with rising drum patterns or risers.
    \item \textbf{Drop:} Peak energy section, featuring main rhythm motifs and basslines.
    \item \textbf{Breakdown:} Reduced energy sections, often melodic or atmospheric, providing contrast to drops.
    \item \textbf{Outro:} Ending portion, with gradually decreasing energy.
    \item \textbf{Silence:} Sections with minimal or no audio signal.
    \item \textbf{End:} Explicit track ending or final note.
\end{itemize}

\begin{figure*}[t]
    \centering
    \includegraphics[width=\textwidth]{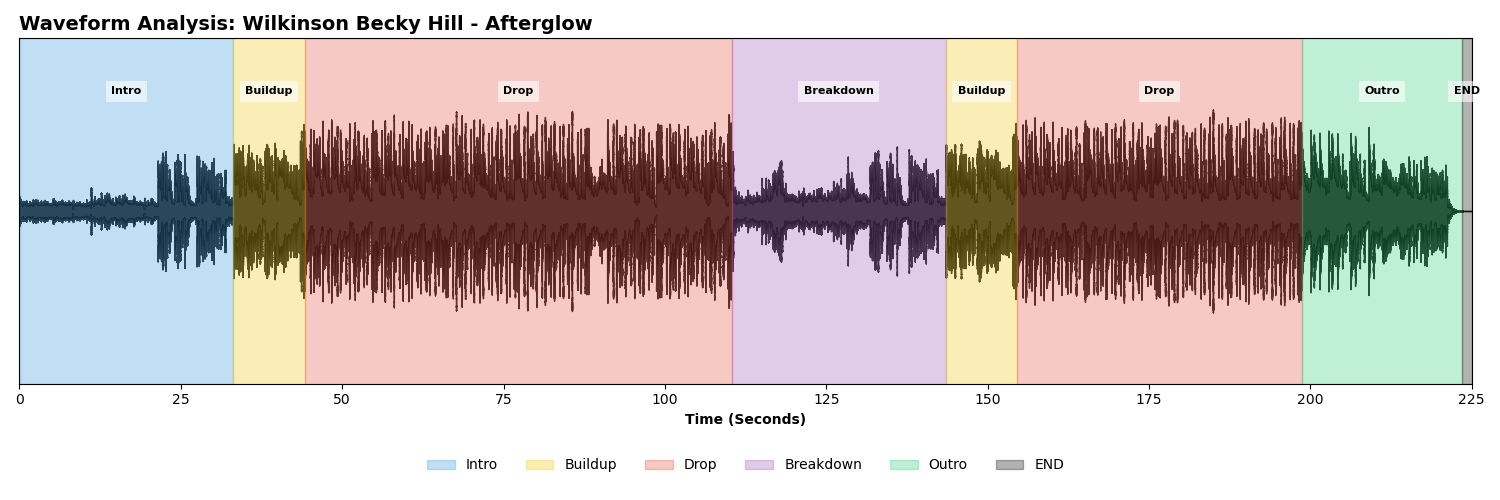}
    \caption{Example track segment annotated with proposed EDM taxonomy.}
    \label{fig:annotations_example}
\end{figure*}

Figure~\ref{fig:annotations_example} illustrates an example track annotated with our proposed taxonomy, highlighting EDM's structural transitions.

\subsection{Model Architecture}

Our model architecture largely followed SongFormer's design, with the key differences being our incorporation of an EDM-specific taxonomy, a training stage using EDM-98, and a final fine-tuning stage on EDM-98 \cite{songformer2025}.

We extracted self-supervised music representations from two pretrained audio foundation models, MuQ and MusicFM  \cite{muq2025, won2023musicfm}. We selected these two as our foundation models because they both produce temporally rich embeddings from relatively large audio corpora, providing strong low-level acoustic representations. Critically, however, neither model's pretraining is optimized for EDM structure. Their learned representations reflect general acoustic patterns rather than the energy and timbral transitions that define EDM sections. This generality is precisely why domain-specific fine-tuning over our dataset EDM-98 is necessary. EDMFormer repurposed these general embeddings using an EDM-specific framework, teaching the model which acoustic patterns are structurally meaningful for the genre.

We fused both models as they differ in their pretraining objectives: MuQ uses masked prediction over Mel-quantized spectral targets, emphasizing timbral and spectral features, while MusicFM uses masked token modeling with random projection and has demonstrated stronger performance on longer-range structural tasks due to its 30-second input context during pretraining \cite{muq2025, won2023musicfm}. Our combination of the two methods provides complementary information to the downstream transformer. Each model produced 1024-dimensional embeddings computed at two temporal contexts (30 second and 420 second windows), resulting in four embedding streams per track. The four embeddings were then concatenated to form a 4096-dimensional representation. Next, this concatenated 4096-dimensional representation was projected to 2048 dimensions through a linear layer and provided as the input to the SongFormer encoder.

\begin{figure*}[t]
    \centering
    \includegraphics[width=\textwidth]{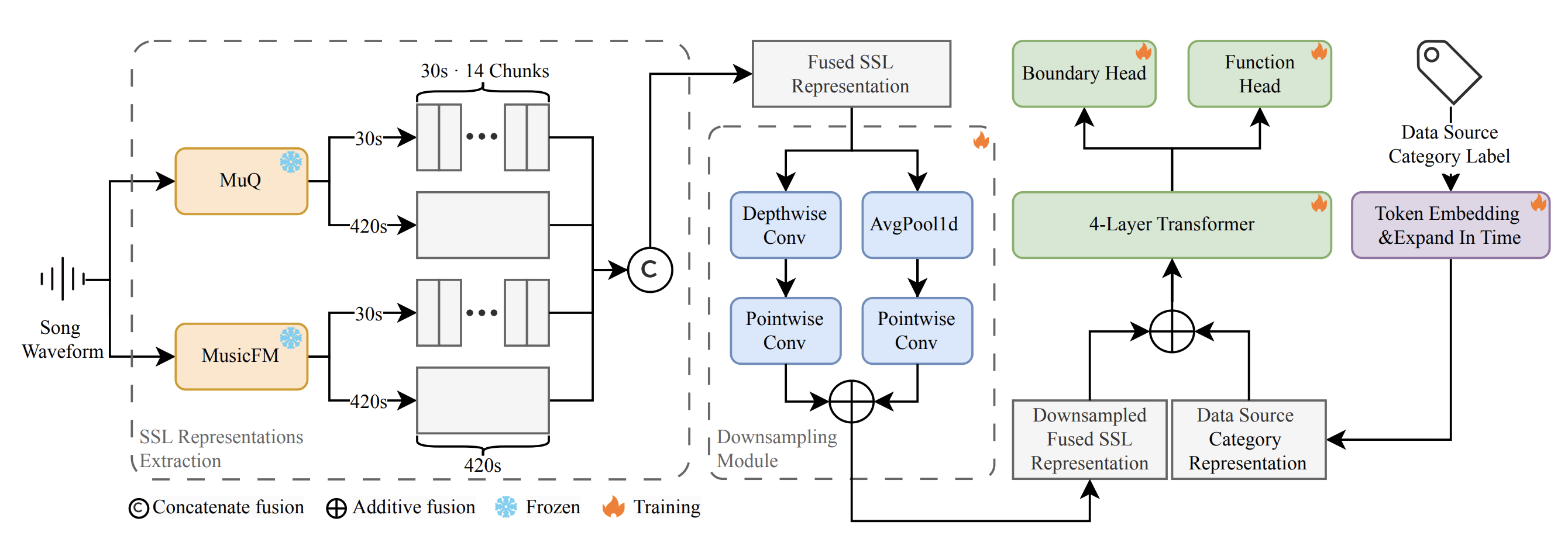}
    \caption{SongFormer's model architecture.}
    \label{fig:model_architecture}
\end{figure*}

Figure~\ref{fig:model_architecture} visualizes SongFormer's model architecture in greater detail for reference.

\subsection{Baseline}

We evaluated EDMFormer against the newest iteration of SongFormer (2025) as our primary baseline \cite{songformer2025}. SongFormer is the current best SSL-based MSA model, having demonstrated strong performance on pop-centric benchmarks such as SALAMI \cite{salami2011}. It is pretrained on diverse music corpora using heterogeneous supervision and uses a general structural taxonomy (e.g. intro, verse, chorus, bridge, outro).

SongFormer serves as an ideal baseline for two reasons. First, it shares the same foundation as EDMFormer, allowing us to isolate the improvements that come with domain-specific data and taxonomy. Second, SongFormer reflects the current standard against which any MSA model should be measured. We evaluated SongFormer in a zero-shot transfer setting, applying it directly to our tracks from EDM-98 without fine-tuning, testing how well general MSA models can analyze EDM.

\subsection{Evaluation}

We evaluated results using hit rate (HR), which measures the percentage of predicted boundaries that fall within a tolerance window of a ground-truth boundary. The metrics in specific were HR@0.5s, HR@3s, and per-frame label accuracy (ACC) with an 88-10 train-test split across the EDM dataset. A 5-fold cross-validation was performed on the training set for hyperparameter tuning.

\section{Results}

Table~\ref{tab:mainresults} compares the performance of our proposed model EDMFormer against the industry's baseline SongFormer (rounded to 3 decimal points).
\begin{table}[H]
\centering
\caption{EDMFormer vs. SongFormer Performance}
\label{tab:mainresults}
\begin{tabular}{|l|c|c|c|}
\hline
\textbf{Model} & \textbf{HR@0.5} & \textbf{HR@3} & \textbf{ACC} \\ \hline
SongFormer (Pop Taxonomy) & 0.569 & 0.608 & 0.148 \\ \hline
EDMFormer (EDM Taxonomy) & 0.616 & 0.635 & 0.883 \\ \hline
\end{tabular}
\end{table}

As shown in Table I, EDMFormer outperformed SongFormer across all three evaluation metrics, with improvements of 4.7\%, 2.7\%, and 73.5\% in HR@0.5, HR@3, and ACC, respectively. The most striking gain was in per-frame accuracy (ACC), where EDMFormer achieved 0.883 compared to SongFormer's 0.148, a 73.5\% improvement that reflects the fundamental mismatch between a pop-centric taxonomy and EDM's structural patterns. Boundary detection improvements further confirm that EDMFormer learns more temporally precise transition cues, such as drop onsets and buildup peaks, which the pop-centric SongFormer systematically misses.

The magnitude of improvement across all three metrics supports the core hypothesis that genre mismatch is a primary driver of poor MSA performance on EDM. EDMFormer's architecture, which fuses MuQ and MusicFM embeddings at two temporal scales before passing the combined representation to the SongFormer encoder, preserved the expressive power of general SSL embeddings while allowing the transformer to specialize its attention patterns toward energy-driven structural transitions. The consistent directional improvement across all metrics and folds indicates that the proposed combination of domain-specific data, taxonomy, and SSL-based architecture is an effective and reproducible approach for adapting music structure analysis to specialized genres.

\section{Conclusion}

This research introduces the EDM-98 dataset, proposes an EDM-specific structural taxonomy, and demonstrates that self-supervised music representations with a transformer model can improve boundary detection and section labelling accuracy for EDM. These results indicate that combining genre-specific datasets and taxonomies with SSL-based transformer models is an effective approach for adapting MSA to specialized genres.

Several limitations constrain the current study. EDM-98 has only 98 tracks. While this amount is sufficient as a proof-of-concept, it represents only a small subset of the broader EDM-CUE corpus, with dataset growth bottlenecked by the time needed for high-precision manual annotation. Another limitation lies in the foundation models themselves: although MuQ and MusicFM provide strong general acoustic representations, their pretraining objectives are not optimized for the energy-driven structural logic of EDM, and may inadequately capture the timbral and dynamic transitions. Lastly, each track in EDM-98 was labeled by a single annotator, potentially introducing bias.

Future work can address these gaps through dataset scaling, integration of explicit acoustic cues, and inter-annotator agreement metrics. Future work might also include hybrid feature learning, hierarchical annotations, relative positional embeddings for long tracks, and real-time inference capabilities for DJ-oriented applications. Implementing these ideas would strengthen both the reliability and representational capacity of future iterations, with longer-term potential for domain-specific SSL pretraining and real-time inference for DJ-oriented applications.

\section{Acknowledgements}

We thank the annotators for their contributions to the development of the EDM-98 dataset and the open-source communities behind SongFormer, MuQ, MusicFM, and librosa. This research was supported through the mentorship and computational resources provided by the University of Waterloo's AI team, WAT.ai. We also thank the Canadian Undergraduate Conference on Artificial Intelligence (CUCAI) for the opportunity to present this work.

\newpage
\bibliographystyle{IEEEtran}
\bibliography{references}

\end{document}